\documentclass[prb,twocolumn,showpacs,superscriptaddress]{revtex4}

\usepackage[dvips]{graphicx}
\usepackage{amsmath}
\usepackage{placeins}

\begin{document}
\title{Topological magnetic order and superconductivity in EuRbFe$_4$As$_4$}

\author{M.~Hemmida}
\thanks{These two authors contributed equally}
\affiliation{Experimental Physics V, Center for Electronic Correlations and Magnetism, University of Augsburg, D-86135 Augsburg, Germany}

\author{N.~Winterhalter-Stocker}
\thanks{These two authors contributed equally}
\affiliation{Experimental Physics VI, Center for Electronic Correlations and Magnetism, University of Augsburg, D-86135 Augsburg, Germany}

\author{D.~Ehlers}
\affiliation{Experimental Physics V, Center for Electronic Correlations and Magnetism, University of Augsburg, D-86135 Augsburg, Germany}

\author{H.-A.~Krug~von~Nidda}
\affiliation{Experimental Physics V, Center for Electronic Correlations and Magnetism, University of Augsburg, D-86135 Augsburg, Germany}

\author{M.~Yao}
\affiliation {Max Planck Institute for Chemical Physics of Solids, D-01187 Dresden, Germany}

\author{J.~Bannies}
\altaffiliation{Present address: Quantum Matter Institute, University of British Columbia, Vancouver, BC V6T 1Z4, Canada}
\affiliation{Max Planck Institute for Chemical Physics of Solids, D-01187 Dresden, Germany}

\author{E.D.L.~Rienks}
\affiliation{Helmholtz-Zentrum Berlin, Albert-Einstein-Strasse 15, D-12489 Berlin, Germany}

\author{R.~Kurleto}
\affiliation {Leibniz Institute for Solid State and Materials Research  Dresden, Helmholtzstr. 20, D-01069 Dresden, Germany}
\affiliation {M. Smoluchowski Institute of Physics, Jagiellonian University, {\L}ojasiewicza 11, 30-348, Krak{\'o}w, Poland }

\author{C.~Felser}
\affiliation {Max Planck Institute for Chemical Physics of Solids, D-01187 Dresden, Germany}

\author{B.~B\"uchner}
\affiliation{Leibniz Institute for Solid State and Materials Research  Dresden, Helmholtzstr. 20, D-01069 Dresden, Germany}
\affiliation {Institut f\"ur Festk\"orperphysik, Technische Universität Dresden, D-01062 Dresden, Germany}

\author{J.~Fink}
\affiliation{Leibniz Institute for Solid State and Materials Research  Dresden, Helmholtzstr. 20, D-01069 Dresden, Germany}
\affiliation {Max Planck Institute for Chemical Physics of Solids, D-01187 Dresden, Germany}
\affiliation {Institut f\"ur Festk\"orperphysik, Technische Universität Dresden, D-01062 Dresden, Germany}

\author{S.~Gorol}
\affiliation{Experimental Physics VI, Center for Electronic Correlations and Magnetism, University of Augsburg, D-86135 Augsburg, Germany}

\author{T.~F\"orster}
\affiliation{Hochfeld-Magnetlabor Dresden (HLD-EMFL) and W\"urzburg-Dresden Cluster of Excellence ct.qmat, Helmholtz-Zentrum Dresden-Rossendorf, D-01328 Dresden, Germany}

\author{S.~Arsenijevic}
\affiliation{Hochfeld-Magnetlabor Dresden (HLD-EMFL) and W\"urzburg-Dresden Cluster of Excellence ct.qmat, Helmholtz-Zentrum Dresden-Rossendorf, D-01328 Dresden, Germany}

\author{V.~Fritsch}
\affiliation{Experimental Physics VI, Center for Electronic Correlations and Magnetism, University of Augsburg, D-86135 Augsburg, Germany}

\author{P.~Gegenwart}
\affiliation{Experimental Physics VI, Center for Electronic Correlations and Magnetism, University of Augsburg, D-86135 Augsburg, Germany}

\date{\today}

\begin{abstract}
 
We study single crystals of the magnetic superconductor EuRbFe$_4$As$_4$ by magnetization, electron spin resonance (ESR), angle-resolved photoemission spectroscopy (ARPES) and electrical resistance in pulsed magnetic fields up to $630$~kOe. The superconducting state below $36.5$~K is almost isotropic and only weakly affected by the development of Eu$^{2+}$ magnetic order at $15$~K. On the other hand, for the external magnetic field applied along the \textit{c}-axis the temperature dependence of the ESR linewidth reveals a Berezinskii-Kosterlitz-Thouless topological transition below $15$~K. This indicates that Eu$^{2+}$-planes are a good realization of a two-dimensional XY-magnet, which reflects the decoupling of the Eu$^{2+}$ magnetic moments from superconducting FeAs-layers.

\end{abstract}


\maketitle


\section{Introduction}

Magnetic superconductors and superconducting magnets are very intriguing materials due to the competition of magnetic order and superconductivity. Theoretical predictions made by Ginzburg showed that uniform magnetism in bulk compounds may destroy superconductivity due to the electromagnetic mechanism.\cite{Ginzburg1956} For example, the incompatible nature of superconductivity and ferromagnetism was demonstrated by experiments, which showed the competition of the two collective phenomena in (La,Gd) and (Ce,Pr)Ru$_2$ solid solutions.\cite{Matthais1958} The suppression of ferromagnetism in the superconducting regime was explained by Anderson and Suhl in terms of the Ruderman-Kittel-Kasuya-Yosida (RKKY) interaction by the end of the 1950s.\cite{Anderson1959} Judging from the energy scale, however, ferromagnetism wins over superconductivity in most cases. Thus, it was suggested that in the superconducting state, the spin susceptibility is suppressed at small wavevectors and pure ferromagnetism should be modified in the form of crypto-ferromagnetic alignment for localized spins.\cite{Anderson1959} Only in the late 1970s the coexistence of superconductivity and ferromagnetism was evidenced in ErRh$_4$B$_4$ \cite{Fertig1977} and Ho$_{1.2}$Mo$_6$S$_8$ \cite{Ishikawa1977} in narrow regimes of temperature and external magnetic field. In the late 1990s, superconductivity and weak ferromagnetism were observed in high-temperature superconductor rutheno-cuprates.\cite{Felner1997,Bernhard1999} In the examples above, superconductivity and ferromagnetism obviously originate from different electrons of different elements. However, there is a scenario that both superconductivity and ferromagnetism arise from the same type of electrons: e.g. in UGe$_2$ \cite{Saxena2000} and URhGe \cite{Aoki2001} the superconductivity emerges from the ferromagnetic background ($T_{\rm c}$ $<$ $T_{\rm m}$), where $T_{\rm m}$ is the magnetic transition temperature. Such compounds are called superconducting magnets, while magnetic superconductors are known for the case of $T_{\rm c}$ $>$ $T_{\rm m}$.

Contrary to bulk materials, the coexistence of superconductivity and ferromagnetism may easily be achieved in artificially fabricated superconductor/ferromagnet heterostructures. Due to the proximity effect the Cooper pairs penetrate into the ferromagnetic layer giving the unique possibility to study properties of superconducting electrons under the influence of the huge exchange field. The proximity effect at superconductor/ferromagnet interfaces produces a damped oscillatory behavior of the Cooper pair wave function within the ferromagnetic medium.\cite{Buzdin2005} In inhomogeneous superconductivity, an analogous effect was predicted a long time ago which is well known as the Fulde-Ferrel-Larkin-Ovchinikov (FFLO) effect.\cite{Larkin1964,Fulde1964} This effect first was suggested for a pure ferromagnetic superconductor at low temperatures. Moreover, by variation of the nanoscale thickness of the ferromagnetic and superconducting layers in a controllable manner it is possible to change the relative strength of the two competing ordering mechanisms.\cite{Zdravkov2006,Lenk2016}

Fe-based superconductors are characterized by multiband superconductivity as well as high transition temperatures. This feature makes it possible to see new phenomena including those due to the interplay of superconductivity and magnetism.\cite{Kamihara2008} Evidence of the coexistence of superconductivity and ferromagnetism was observed, for example, in SrFe$_2$As$_2$~\cite{Saha2009} due to the lattice distortions. Coexistence of superconductivity and ferromagnetism was also observed in other iron-pnictide systems like Sr$_2$VFeAsO$_3$~\cite{Cao2010} and CeFe(As$_{\rm 1-x}$P$_{\rm x}$)O$_{0.95}$F$_{0.05}$~\cite{Luo2011} were it results from Vanadium and Ce ions, respectively. 

An outstanding example comes from Eu-based iron pnictides, especially EuFe$_2$As$_2$-related systems, in which the Eu$^{2+}$ ions show large local magnetic moments with $J = S = 7/2$. The Eu$^{2+}$ magnetic moments in EuFe$_2$As$_2$ are coupled ferromagnetically within the \textit{ab}-planes, but antiferromagnetically along the \textit{c}-axis. It means that Eu$^{2+}$ magnetic moments are rotated by $180^{\circ}$ from plane to plane.\cite{Xiao2009} The compound undergoes a spin-density-wave (SDW) order in the Fe sublattice accompanied by a tetragonal-to-orthorhombic structural phase transition below $T_{\rm SDW}=195$~K.\cite{Jeevan2008} Partial substitution of Fe by Ru~\cite{Jiao2011,Jiao2012} or Ni~\cite{Ren2009a} in EuFe$_2$As$_2$ suppresses the SDW transition. This process is accompanied by the appearance or absence of superconductivity for Ru and Ni doping, respectively. Both cases are associated with the emergence of ferromagnetic ordering of Eu$^{2+}$ magnetic moments. Ferromagnetic ordering of Eu$^{2+}$ magnetic moment in EuFe$_2$As$_2$ was also achieved by the partial substitution of As by the isoelectronic P.\cite{Nandi2014} It was found that with increasing P substitution, the Eu$^{2+}$ magnetic moments cant out of the \textit{ab}-plane, yielding a net ferromagnetic component along the \textit{c}-direction. The coexistence of superconductivity and ferromagnetism induced by chemical substitution was observed and confirmed by various methods.\cite{Ren2009,Jiang2009,Ahmed2010,Jiao2011,Nowik2011,Wu2011,Jeevan2011,Zapf2011,Munevar2014,Hemmida2014} 

Very recently, new members of the iron-pnictide family, the so-called $1144$-system $A$$B$Fe$_4$As$_4$ ($A$ = Ca, Sr, Ba, Eu; $B$ = K, Rb, Cs) realize the coexistence of ferromagnetism and superconductivity.\cite{Iyo2016,Kawashima2016,Liu2016,Wang2017,Bao2018} The $1144$ systems can be viewed as $50\%$ hole doped $122$ iron pnictides with ordered stacking of $A^{2+}$ and $B^{1+}$ separating the FeAs layers. In EuRbFe$_4$As$_4$, the Eu$^{2+}$ magnetic moments align ferromagnetically within the \textit{ab}-planes, but rotate by $90^{\circ}$ from plane to plane along the \textit{c}-axis.\cite{Iida2019} On the other hand, EuRbFe$_4$As$_4$ undergoes a superconducting transition above the magnetic one ($T_{\rm c}$ $>$ $T_{\rm m}$). These findings motivated intensive theoretical\cite{Devizorova2019a,Devizorova2019b,Koshelev2019a,Xu2019,Nejadsattari2020} and experimental\cite{Liu2017,Jackson2018,Smylie2018,Stolyarov2018,Smylie2019,Xiang2019,Willa2019,Iida2019,Koshelev2019,Holenstein2019,Liu2020,Willa2020,Vlasov2020,Vlasenko2020,Kim2020} works in order to understand the interplay between these two antagonistic phenomena.

In this comprehensive study, we report synthesis of single crystalline samples of EuRbFe$_4$As$_4$ and their magnetic and transport characterizations. Also we outline experimental details of ESR, resistivity at high-magnetic fields and ARPES. The analysis of ESR data shows that the spin dynamics of Eu$^{2+}$ ions is ascribed to the Berezinskii-Kosterlitz-Thouless scenario. On the other hand, the analysis of upper critical field data reveals an almost isotropic superconductivity. The non- or rather the weak interaction between conduction electrons of FeAs-layers and localized Eu$^{2+}$ magnetic moments is also discussed in the frame of ESR and ARPES results.

\section{Experimental Details}

Single crystals of EuRbFe$_4$As$_4$ were grown using FeAs flux with the same method described by Meier \textit{et al.} for the synthesis of CaKFe$_4$As$_4$.\cite{Meier2017} Via mechanical cleaving the crystals can be removed out of the matrix of FeAs flux and potential RbFe$_2$As$_2$ and EuFe$_2$As$_2$ foreign phases can be eliminated. With this method very thin crystal plates can be extracted with lateral dimension up to 2\,mm x 4\,mm. The crystal plate equates the \textit{ab}-plane and the tetragonal \textit{c}-axis is perpendicular to this plane.

Magnetization measurements have been performed using a commercial magnetometer (Quantum Design MPMS3) at temperatures $2 \leq T \leq 300$~K and in external magnetic field of $10$~Oe. Samples have been measured on heating following the zero-field-cooled (ZFC) as well as field-cooled (FC) measurement protocol. 

The resistivity was measured on single crystals in steady magnetic fields up to $140$~kOe for $2 \leq T \leq 300$~K using a physical properties measurement system (Quantum Design PPMS) with the electrical transport option ($\nu=117$~Hz). For these measurements plate-like crystals with lateral dimensions up to 2\,mm x 4\,mm were employed. Utilizing a four point probe the sample was connected via silver epoxy to Pt wires. Furthermore, measurements at high magnetic fields up to $630$~kOe were carried out using a nondestructive pulsed-field coil at the Dresden High Magnetic Field Laboratory.

ESR measurements were performed in a continuous wave spectrometer (Bruker ELEXSYS E500) at X- and Q-band frequency ($\nu \approx 9.35$ and $34$~GHz, respectively) in the temperature region $4 \leq T \leq 300$~K using a continuous He gas-flow cryostat (Oxford Instruments). ESR detects the power \textit{P} absorbed by the sample from the transverse magnetic microwave field as a function of the static magnetic field \textit{H}. The signal-to-noise ratio of the spectra is improved by recording the derivative $dP/dH$ using a lock-in technique with field modulation. 

ARPES measurements were conducted at the $1^3$-ARPES end station attached to the beamline UE112 PGM at BESSY, equipped with a Scienta R4000 energy analyzer. All data presented in this contribution were taken at temperatures between $1$ and $50$~K. The achieved energy and angle resolutions were between $4$ and $10$~meV and 0.2$^\circ$, respectively. Polarized photons with energies $h\nu=20-130$~eV were employed to reach different $k_z$ values in the BZ and spectral weight with a specific orbital character.\cite{Fink2009,Moser2017} Inner potentials between of $12$ and $15$~eV were used to calculate the $k_z$ values from the photon energy.

\section{Experimental Results and Discussion}
\subsection{Structure and Magnetic Characterizations}

\begin{figure}
\centering
\includegraphics[width=70mm,clip]{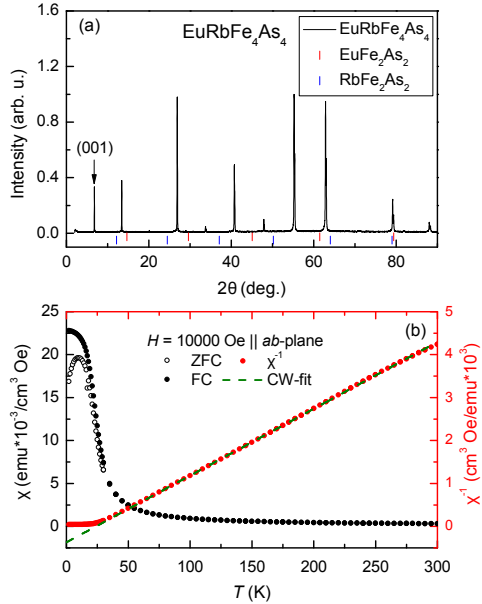}
\caption{(a) Diffractogram of EuRbFe$_4$As$_4$ with the marked peak positions of RbFe$_2$As$_2$ and EuFe$_2$As$_2$. A clear peak at low angle marks the characteristic (001) reflection. (b) Temperature dependence of the magnetic susceptibility measured in a field of $10000$~Oe for $H || ab$. The open symbols show the ZFC measurement and the closed symbols the FC measurement. The red data points correspond to the red axis on the right side and show $1/\chi$. This data show the Curie-Weiss like behavior at high temperatures with a Curie-Weiss temperature of $\Theta_{\rm CW} = 24.7$~K and an effective moment of $\mu_{\rm eff} \approx 7.98\mu_{\rm B}/f.u.$.}
\label{xrd}
\end{figure}

The sample quality has been confirmed by means of x-ray diffraction (XRD) analysis. In Fig.~\ref{xrd}(a) the room-temperature XRD pattern is shown. The presence of the \textit{h+k+l=odd} peaks indicates the ordered \textit{P/4mmm} structure, because these peaks would be forbidden in the \textit{I4/mmm} order of the 122-structure. There is no visible signature of EuFe$_2$As$_2$ and RbFe$_2$As$_2$ (00\textit{l}) peaks which are the most common impurity phases. The peak positions are in good agreement with the lattice constants reported by Bao \textit{et al.}~\cite{Bao2018} ($a=0.38825$~nm, $c=1.32733$~nm).

In Fig.~\ref{xrd}(b) the temperature dependence of the magnetic susceptibility $\chi=M/H$ for $H || ab$ at $10000$~Oe is shown. The compound is dominated by the Curie paramagnetic contribution of the localized Eu$^{2+}$ moments ($J = S = 7/2$). The positive value of Curie-Weiss temperature $\Theta_{\rm CW} \approx 25$~K indicates the predominant ferromagnetic nature of the exchange interaction. The effective moment was determined as a mean value of the effective moments of different directions to be $\mu_{\rm eff} = 7.98\mu_{\rm B}/f.u.$. This value is close to the theoretical value $\mu_{\rm eff} = g\mu_{\rm B}\sqrt{J(J+1)} \approx 7.94\mu_{\rm B}/f.u.$, which confirms the $2+$ state of Europium in EuRbFe$_4$As$_4$. 

\begin{figure}
\centering
\includegraphics[width=70mm,clip]{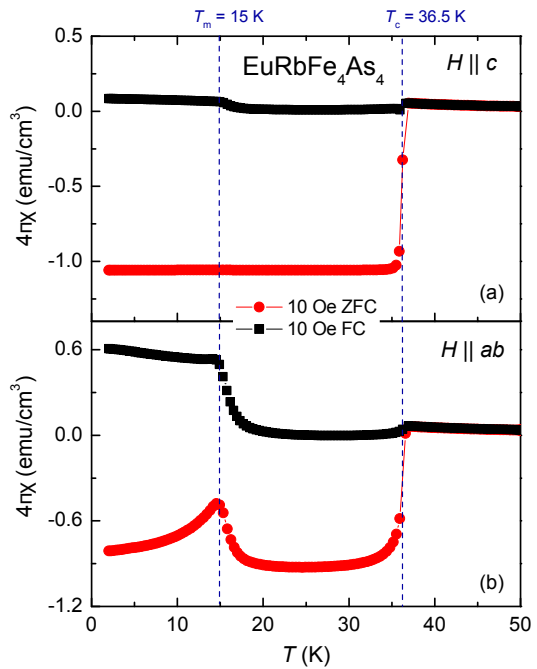}
\caption{Temperature dependence of magnetic susceptibility $\chi = M/H$ of EuRbFe$_4$As$_4$ in a field of $10$~Oe applied along the \textit{c}-axis and the \textit{ab}-plane as shown in (a) and (b), respectively. The red data points correspond to the \textit{ZFC} measurement and the black ones to the \textit{FC} measurements. From these measurements a superconducting transition temperature $T_{\rm c} = 36.5$~K can be determined. The Eu$^{2+}$ magnetic moments order ferromagnetically at $T_{\rm m} = 15$~K.}
\label{magnetization}
\end{figure}

A field of $10$~Oe was applied along the \textit{c}-axis and within the \textit{ab}-plane (see Fig.~\ref{magnetization}(a,b)). At $36.5$~K a sharp downturn marks the superconducting phase transition. The diamagnetic signal for $H || c$ is close to $4\pi\chi = -1$ and indicates a complete superconducting volume. At $T_{\rm m} = 15$~K a kink-like anomaly in the ZFC measurement indicates the ordering of Eu$^{2+}$ within the superconducting state. In contrast to $H || ab$, the magnetic signal at $T_{\rm m} = 15$~K is only marginal. This is an indication that Eu$^{2+}$ moments prefer the \textit{ab}-plane as an easy-plane and hence represent a good realization of the two-dimensional XY-model. In the \textit{ab}-plane the FC measurement shows only a small kink which indicates the transition. This is due to the fact that in the FC condition the flux is frozen in the sample and no complete Meissner-state can be established.   
 
\begin{figure}
\centering
\includegraphics[width=70mm,clip]{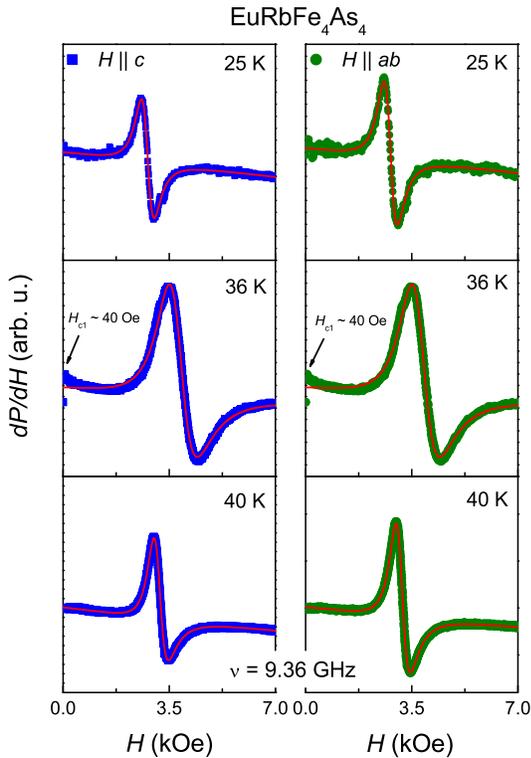}
\caption{ESR spectra in X-band ($\nu \approx 9.35$~GHz) for selected temperatures in the paramagnetic regime of EuRbFe$_4$As$_4$ along and perpendicular to the $c$-axis below, near and above $T_{\rm c} = 36.5$~K in the paramagnetic regime. The solid line indicates the fit with the field derivative of an asymmetric Lorentz line. A front peak at $36$~K around $H_{\rm c1}=40$~Oe exhibits the onset of superconducting state for both directions.}
\label{Spectra}
\end{figure}

\begin{figure}
\centering
\includegraphics[width=75mm,clip]{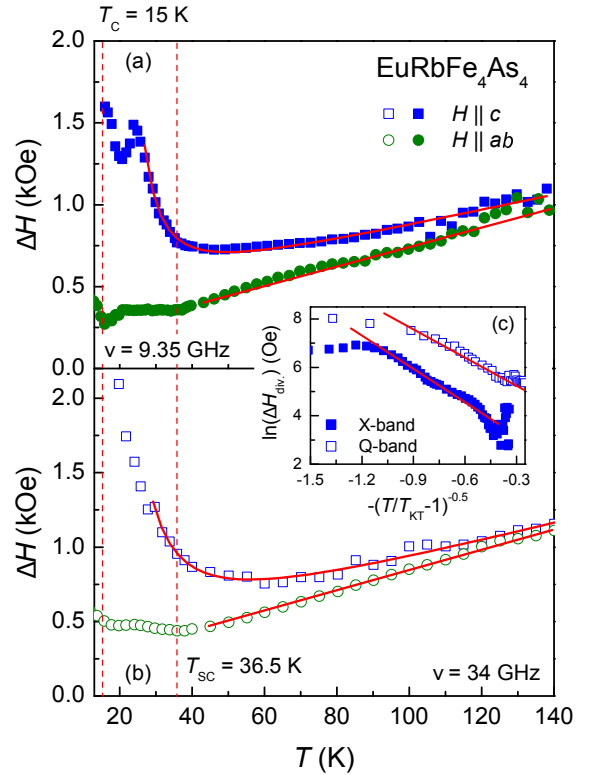}
\caption{In captions (a) and (b) temperature dependence of the ESR linewidth of EuRbFe$_4$As$_4$ measured at X- and Q-bands ($\nu = 9.35$ and $34$~GHz, respectively). Solid lines represent a combined formula of Korringa and BKT formulae for $H || c$: $Eq.~\ref{Korringa}+Eq.~\ref{BKT}+\Delta H_{\rm 0}$ where $\Delta H_{\rm 0}$ is the residual linewidth. The inset (c) shows the quality of the BKT fit using logarithmic plot ln($\Delta H_{\rm div}$) vs the reduced temperature $-(T/T_{\rm KT}-1)^{-0.5}$; $\Delta H_{\rm div}= Eq.~\ref{BKT}-Eq.~\ref{Korringa}-\Delta H_{\rm 0}$. For $H || ab$ only linear Korringa fit is applied for $T>40$~K.}
\label{Linewidth}
\end{figure}

Figure~\ref{Spectra} shows ESR spectra of EuRbFe$_4$As$_4$ below, near and above $T_{\rm c} = 36.5$~K in the paramagnetic regime for the magnetic field aligned along different crystallographic directions. All spectra in this regime exhibit a single exchange-narrowed resonance which is well described by an asymmetrical Lorentz line due to the skin effect. The skin effect appears in metals because of the interaction between the applied microwave field and mobile charge carriers. This leads to an admixture of dispersion $\chi^{'}$ to the absorption $\chi^{''}$ depending on the ratio of skin depth and sample size.\cite{Barnes1981} The ratio  $\chi^{'}/\chi^{''}$ is found to change from $0.1$ above $T_{\rm m}$ to values slightly larger than $1$ above $T_{\rm c}$ (paramagnetic metal). 

In the vicinity of $T_{\rm c}$, the ESR spectra show non-resonant absorption (front peaks or bumps) in small fields below $50$~Oe for both directions as signatures of the onset of the superconducting state (see also Fig.~\ref{critical}(b)). The front peak is due to the surface resistivity which changes as a function of the magnetic field.\cite{Owens2001} These front peaks remain nearly unchanged in the temperature domain of $3-5$~K. As the temperature increases and the system reaches $T_{\rm c}$ the front peak disappears and the system becomes a conventional paramagnetic metal. As the linewidth $\Delta H$ is large enough and comparable to the order of magnitude of the resonance field $H_{\rm res}$, the counter-resonance at $-H_{\rm res}$ had to be considered in the fit.\cite{Joshi2004} The \textit{g}-factor at high temperatures is close to $2$ for both $H || ab$ and $H || c$. The resonance field shifts to lower fields on approaching magnetic order on account of the local ferromagnetic polarization, while the $g$-value increases. However, taking into account the demagnetization factor the corrected $g$-factor near $4$~K is estimated as $g_{\rm ab} = 2.068$ and $g_{\rm c} = 2.023$. 

The most important information is obtained from the temperature dependence of the linewidth (Fig.~\ref{Linewidth}). The linewidth $\Delta H$ strongly increases upon approaching the magnetic transition $T_{\rm m} = 15$~K from above. On the other hand $\Delta H$ starts to increase linearly with temperature above $T_{\rm c} = 36.5$~K as well. This indicates the dominant role of the Korringa relaxation of the localized Eu$^{2+}$ spins via scattering of the conduction electrons:
\begin{equation}
\Delta H = \frac{\pi k_{\rm B}}{g\mu_{\rm B}}\left\langle J^{2}(q) \right\rangle D^{2}(E_{\rm F})T = mT
\label{Korringa}
\end{equation}
where $\left\langle J^{2}(q) \right\rangle$ is the squared exchange constatnt between localized spins and conduction electrons averaged over the momentum transfer $q$, $D(E_{\rm F})$ is the conduction-electron density of states at Fermi energy $E_{\rm F}$ and $m$ is the Korringa slope.\cite{Korringa1950,Barnes1981} A typical value of $m$ in Eu-based iron pnictides is about $8$~Oe/K.\cite{Hemmida2014,Dengler2010,Pascher2010,Krug2012} This value is typical for the $S$-state of $4f^{7}$ local moments in conventional metals as well.\cite{Korringa1950,Barnes1981,Taylor1975} 

According to Willa \textit{et al.} in Ref.~\onlinecite{Willa2019}, specific heat measurements under magnetic field along the \textit{c}-axis up to $3$~kOe reveal a topological phase transition--the so-called Kosterlitz-Thouless phase transition--at $T_{\rm KT} \approx 9$~K. This finding was confirmed by Monte-Carlo simulations of an easy-plane two-dimensional Heisenberg model. Following these results, one can apply a Berezinskii-Kosterlitz-Thouless (BKT) scenario\cite{Berezinskii1972,Kosterlitz1973,Kosterlitz1974} in order to describe the relaxation mechanism of the ESR linewidth at low temperatures ($T<60$~K) for $H || c$. It implies that:
\begin{equation}
\Delta H = \zeta^{3} = \Delta H_{\infty}exp[3b/\sqrt{\frac{T}{T_{\rm KT}}-1}]\label{BKT}
\end{equation}
where $\zeta$ is the correlation length of vortices above $T_{\rm KT}$, $\Delta H_{\infty}$ is the ESR linewidth in high-temperature approximation (neglecting any Korringa relaxation) and $b=\pi/2$ for the square lattice (see e.g. Ref.~\onlinecite{Heinrich2003}).

One obtains $T_{\rm KT}\approx 14$~K and $11$~K for X- and Q-band respectively which are in fair agreement with values reported in Ref.~\onlinecite{Willa2019}. The value of $T_{\rm KT}$ is always below the magnetic ordering temperature $T_{\rm m}$, observed in zero-field ($T_{\rm KT}/T_{\rm m}\approx 0.7-0.9$), as typically found in quasi two-dimensional magnets.\cite{Demokritov1989,Gaveau1991,Heinrich2003} Note that in the crossover regime interference between magnetic vortices and three-dimensional ordering fluctuations masks the pure BKT-scenario. Therefore the model fails to describe the domain close $T_{\rm m}$ (see Fig.~\ref{Linewidth}). The Korringa slope $m = 6.0(5)$~Oe/K was found to be nearly isotropic and independent from frequency within the error bars. Similar values of Korringa slope have been reported in other Eu-based FeAs superconductors.\cite{Pascher2010,Krug2012,Hemmida2014}  

Thus, as a conclusion of ESR measurements, the BKT transition at low temperatures proves the two-dimensionality of Eu-magnetism decoupled from the conduction electrons of the FeAs-layers, while the Koringa behavior for high temperatures signifies the three-dimensionality of the metallic phase.

\subsection{Resistivity and Critical Fields}

\begin{figure}
\centering
\includegraphics[width=70mm,clip]{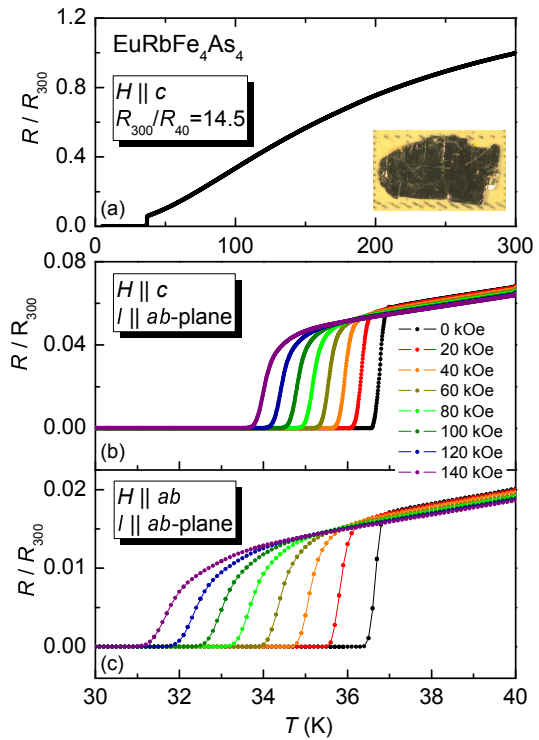}
\caption{(a) Normalized in-plane resistance (\textit{I} $||$ \textit{ab}-plane). The measurement shows a convex curvature. The inset depicts the single crystal under investigation. (b) Shift of the superconducting transition in fields up to $140$~kOe applied along the \textit{c}-axis. (c) Shift of the transition in fields up to $140$~kOe with the field applied within the \textit{ab}-plane.}
\label{resistivity}
\end{figure}

The temperature dependence of the normalized electrical resistivity is shown in Fig.~\ref{resistivity}. Even though the material is metallic, as other $1144$-type and $122$-type superconductors, it shows a convex curvature instead of a normal linear metallic behavior. This feature is associated with multiband effects in hole doped materials where carriers in different bands show different mobilities for different temperatures.\cite{Golubov2011,Shen2011} The residual resistivity ratio (RRR) was determined as $R_{300 K}/R_{40 K} = 14.5$ and hence indicates good crystal quality. No finite resistance occurs at or below the temperature of the magnetic order. The inset depicts the single crystal under investigation. This value is similar to the value reported in Ref.~\onlinecite{Smylie2018}. A sharp superconducting transition occurs at $T_{\rm c} = 36.5$~K. The superconducting transition temperature window is only about $0.4$~K. No reentrance behavior is observable at the ordering temperature of the Eu$^{2+}$ magnetic moments at $T_{\rm m} = 15$~K, as a return to the normal state at this temperature. This shows the unique behavior of this compound due to the strong decoupling of the magnetic and superconducting sublattices. This is in contrast to some Eu containing $122$-iron-based superconductors which show a reentrance behavior.\cite{Jiao2011,Kurita2011,Paramanik2013} These findings are in a good agreement with those in Ref.~\onlinecite{Smylie2018}.

\begin{figure}
\centering
\includegraphics[width=70mm,clip]{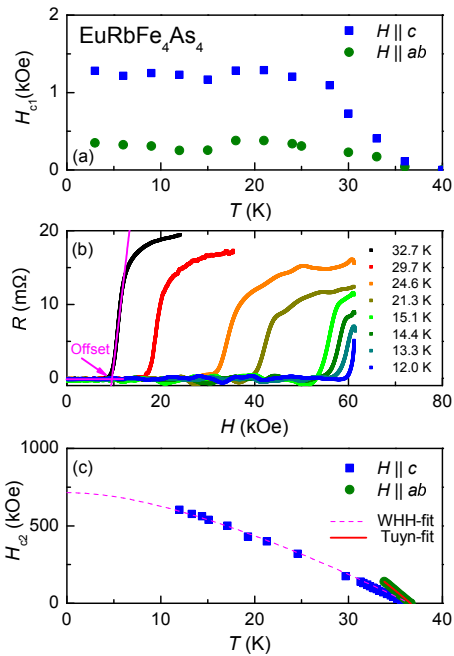}
\caption{(a) The phase diagram of the lower critical field applied along the \textit{c}-axis and in the \textit{ab}-plane. The lower critical field shows a distinct minimum at roughly $15$~K which corresponds to the magnetic ordering temperature. (b) Field-dependent resistivity measurements carried out by pulsed magnet for fields $H || c$ up to $630$~kOe at different temperatures. The arrow marks the offset of the superconducting transition, used to determine $T_{\rm c}$. (c) Phase diagram of the upper critical field of EuRbFe$_4$As$_4$. The data for $H || c$ can be fitted using the Werthamer-Helfand-Hohenberg (WHH) model and indicate a $H_{\rm c2}(0)$ of approximately $730$~kOe. On the other hand $H || ab$ data are well described by using Tuyn's law $H_{\rm c2}(T) \approx H_{\rm c2}(0)[1-(\textit{T}/\textit{T}{\rm c})^{2}]$.}
\label{critical}
\end{figure}

In order to study the superconducting anisotropy, the superconducting transition was studied in various fields applied within the \textit{ab}-plane and along the \textit{c}-axis. The shift of the transition in magnetic fields up to $140$~kOe applied along the \textit{c}-axis and in the \textit{ab}-plane are depicted in Figs.~\ref{resistivity}(a) and (b). In both measurements the current was applied within the \textit{ab}-plane. The superconducting transition is only suppressed by a few Kelvin in a field of $140$~kOe.  As an example for $H || c$ the transition shifts to roughly $34$~K. The suppression is more efficient if the field is applied perpendicular to the \textit{c}-axis but the anisotropy is rather small. For both directions a large slope of the upper critical field $H_{\rm c2}(T)$ was observed (see Fig.~\ref{critical}c) as well as a negative magnetoresistance in the normal conducting region. The origin of the negative magnetoresistance is due to a suppression of the electron scattering by spin fluctuations. A similar behavior was found in EuFe$_2$As$_2$.\cite{Jiang2009a,Terashima2010}

The lower critical field $H_{\rm c1}$ was determined using magnetization measurements for various temperatures. The corresponding phase diagram is shown in Fig.~\ref{critical}(a). At lower temperatures, $H_{\rm c1}$ decreases roughly around $20$~K and shrinks to its minimum value at about $T_{\rm m} = 15$~K in both crystallographic directions. The value of $H^{ab}_{\rm c1}=40$~Oe at $36$~K is equal to that observed in the corresponding ESR spectrum (see Fig.~\ref{Spectra}). However, $H^{c}_{\rm c1}=112$~Oe is larger than the one observed by the ESR.

For a more detailed picture of the superconducting phase diagram, the upper critical field $H_{\rm c2}(T)$ for $H || c$ was determined using field-dependent resistivtiy measurements shown in Fig.~\ref{critical}(b). These measurements were performed at different temperatures in a pulsed magnet up to $630$~kOe. The upper critical field $H_{\rm c2}(T)$ data of $H || ab$ were obtained using PPMS for magnetic fields up to $140$~kOe. 

In Fig.~\ref{critical}(c) the phase diagram for the upper critical field in both directions is shown. For $H || c$, the data exhibit a concave curvature which is well described by Werthamer-Helfand-Hohenberg model~\cite{Werthamer1966} (see Fig.~\ref{critical}(c)). According to Tinkham~\cite{Tinkham1996}, one can estimate the zero temperature coherence length as $\xi^{ab} (0) = [\Phi_{0}/2\pi H^{c}_{\rm c2}(0)]^{0.5}$ where $\Phi_{0}$ is the magnetic flux quantum. 

For the extrapolated upper critical field value at zero temperature $H^{c}_{\rm c2}(0) \approx 730$~kOe, the zero temperature coherence length $\xi^{ab}(0) \approx 2.12$~nm. In order to obtain $H^{ab}_{\rm c2}(T)$, one can use Tuyn's law~\cite{Tuyn1925} $H_{\rm c2}(T) \approx H_{\rm c2}(0)[1-(\textit{T}/\textit{T}{\rm c})^{2}]$ to fit the data as shown in Fig.~\ref{critical}(c). Hence the extrapolated value of $H^{ab}_{\rm c2}(0) \approx 960$~kOe. The extrapolated values of $H^{ab,c}_{\rm c2}(0)$ are nearly identical to the values found in the sister compound CaKFe$_4$As$_4$.\cite{Meier2016} For $\xi^{c}(0) = \Phi_{0}/2\pi H^{ab}_{\rm c2}(0)\xi^{ab}(0) \approx 1.62$~nm. As $\xi^{c}(0)$ is larger than the thickness of the superconducting layer $d = c/2 \approx 0.66$~nm, this indicates that superconductivity in this compound does not split into superconductivity of individual FeAs-layers and hence possesses a three-dimensional and not a two-dimensional character.\cite{Izyumov2010}

Both values of $\xi$ are larger than those found in the same compound in Ref.~\onlinecite{Smylie2018}. The anisotropy factor $\gamma = H^{ab}_{\rm c2}(0)/H^{c}_{\rm c2}(0)$ is found to be $1.32$. This value is very close to that found in single crystalline of Ba$_{0.6}$K$_{0.4}$Fe$_{2}$As$_{2}$.\cite{Yuan2009} It implies that EuRbFe$_4$As$_4$ is a nearly isotropic superconductor as many compounds of $122$-type and appears to be a three-dimensional superconductor. This is a consequence of its Fermi-surface topology. However, several compounds of $1111$-type show considerable amount of anisotropy~\cite{Zhang2011} as in high-temperature cuprates~\cite{Sebastian2008} and organic superconductors\cite{Goddard2004}, which their Fermi surfaces are rather two-dimensional.

Previous high-field measurements in pristine single crystals of EuRbFe$_4$As$_4$~\cite{Smylie2019} reported no crossing of $H^{ab}_{\rm c2}$ and $H^{c}_{\rm c2}$ for $0 < H < 630$~kOe, although it was not ruled out for higher fields $H > 700$~kOe. Indeed in proton-irradiated single crystals of EuRbFe$_4$As$_4$, Smylie et \textit{al.}~\cite{Smylie2019} found that the curves for $H || c$ and $H || ab$ tend to get closer at low temperatures and intersect at $T \approx 10$~K. This indicates that the superconductivity anisotropy becomes reversed.\cite{Smylie2019} The inversion of anisotropy has also earlier been reported in several compounds of Fe-based superconductors such as binary chalcogenides Fe$_{y}$(S,Se)$_{\rm 1-x}$Te$_{x}$.\cite{Lei2010,Fang2010,Khim2010} 

Furthermore, one can roughly estimate the ratio of $\xi^{ab}$ to the mean-free path $l$ using a single-band anisotropic Drude formula $l = \hbar(3n\pi^{2}\sqrt{\epsilon})^{1/3}/n\rho_{n}e^{2}$ where $\rho_{n} \approx 20$~$\mu\Omega$.cm at $T = T_{\rm c}$, $n = 1.25 \times 10^{21}$~cm$^{-3}$, and $\sqrt{\epsilon} = 1/\gamma \approx 0.76$. One obtains $l \approx 50$~nm which is much larger than $\xi^{ab}(0)$ satisfying the clean limit condition.

Now we consider the effects of both Pauli paramagnetism and spin-orbit scattering in a weakly coupled superconductor. In the Pauli paramagnetic limiting case, the so-called Chandrasekhar-Clogston limit is determined by the superconducting energy gap $\Delta$ as~\cite{Chandrasekhar1962,Clogston1962}: 
\begin{equation}
H^{P}(0) = \Delta / \sqrt{g}\mu_{\rm B} = 1.76k_{\rm B}T_{\rm c}/\sqrt{g}\mu_{\rm B}\label{CC} 
\end{equation}
where $g=2$ is the Land\'e factor for a free electron. In the case of multiband scenario by using the value of a narrow Drude gap $\Delta \approx 1.59k_{\rm B}T_{\rm c}$ ($5$~meV) at $4$~K given in Ref.~\onlinecite{Stolyarov2018}, which is close to the estimated value of the middle hole pocket (see Fig.~\ref{DvsT}). It leads to $H^{P}(0) \approx 610$~kOe. On the other hand, the orbital limit of the upper critical field is derived from the WHH theory as~\cite{Werthamer1966}:
\begin{equation}
H^{orb}_{\rm c2}(0) = -0.69\textit{T}_{\rm c}[dH_{\rm c2}(\textit{T})/d\textit{T}]_{\rm \textit{T}=\textit{T}_{\rm c}}\label{WHH}
\end{equation}
The gradient values $[dH^{ab,c}_{\rm c2}(\textit{T})/d\textit{T}]_{\rm \textit{T}=\textit{T}_{\rm c}}$ are found to be $-45$~kOe/K and $20$~kOe/K, which are in the same order of the value found in polycrystalline EuRbFe$_4$As$_4$.\cite{Kawashima2016} Thus it yields $H^{orb}_{\rm c2}(0) \approx 1130$~kOe and $500$~kOe for $H || ab$ and $H || c$, respectively. The value of $H^{orb}_{\rm c2}(0)$ in the \textit{ab}-plane predominates that calculated for $H^{P}(0)$ by a factor $\alpha \approx 2.6$ (compare with $\alpha$ values of several iron-based superconductors given in Ref.~\onlinecite{Zhang2011}), where $\alpha$ is the Maki parameter and represents the relative strength of orbital and spin pair breaking as $\alpha = \sqrt{2}H^{orb}_{\rm c2}(0)/H^{P}(0)$.\cite{Maki1966} For $H || c$, $\alpha \approx 1.2$ is slightly larger than the value of the single-band FFLO instability threshold ($\alpha\approx1$).\cite{Larkin1964} Thus, the pair breaking effect of the magnetic field is more dominated by orbital effects than by the Pauli limit for $H || ab$, while $H^{P}(0)$ exceeds $H^{orb}_{\rm c2}(0)$ along the \textit{c}-axis and therefore the paramagnetic limited effect should become dominant in the characterization of the actual upper critical field.\cite{Kida2009}

As a conclusion of these calculations, we see that the upper critical field in EuRbFe$_4$As$_4$ is nearly isotropic. These findings agree with those results found generally in iron-based superconductors. Furthermore, the spin-paramagnetic effect is the dominant pair-breaking mechanism along the crystallographic \textit{c}-axis, while the spin-orbit effect dominates in the \textit{ab}-plane.

\subsection{ARPES}

\begin{figure}
\centering
\includegraphics [width=7cm]{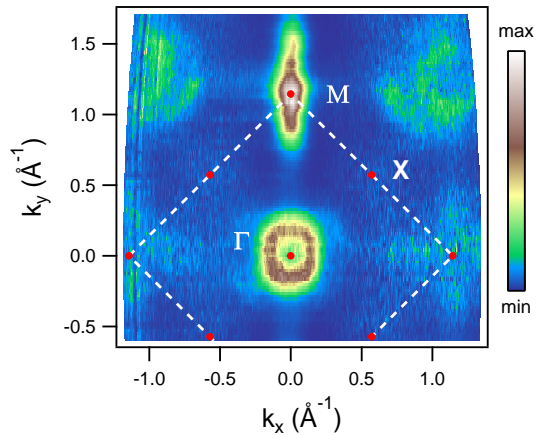}
\caption{ARPES Fermi surface map of EuRbFe$_4$As$_4$ obtained by integrating the photoemission intensity in a $15$~meV window centered at $E_F$. The data were measured at a temperature of $20$~K using vertically polarized photons with an energy of $87$~eV.}
\label{Fermi}
\end{figure}

In Fig.~\ref{Fermi} we present the Fermi surface (FS) of EuRbFe$_4$As$_4$, measured by ARPES~\cite{Damascelli2003} using vertically polarized photons with an energy h$\nu= 87$~eV. The map  shows the FS of the inner hole pocket near the $\Gamma$ point situated at $k_{x,y}=(0,0)$. Due to matrix element effects~\cite{Fink2009,Moser2017} for this photon polarization, the middle hole and the outer hole pocket are not visible at the $\Gamma$ point. On the other hand in the second Brillouin zone at the $\Gamma$ point, traces of both, the middle and the inner hole pockets are detected (see upper right and left corner of the figure. Near $k_{x,y}=(0,1.2)~${\AA}$^{-1}$ at the M point,  the propellerlike electron pocket is visible for this photon polarization.

\begin{figure}
\centering
\includegraphics [width=7cm,clip]{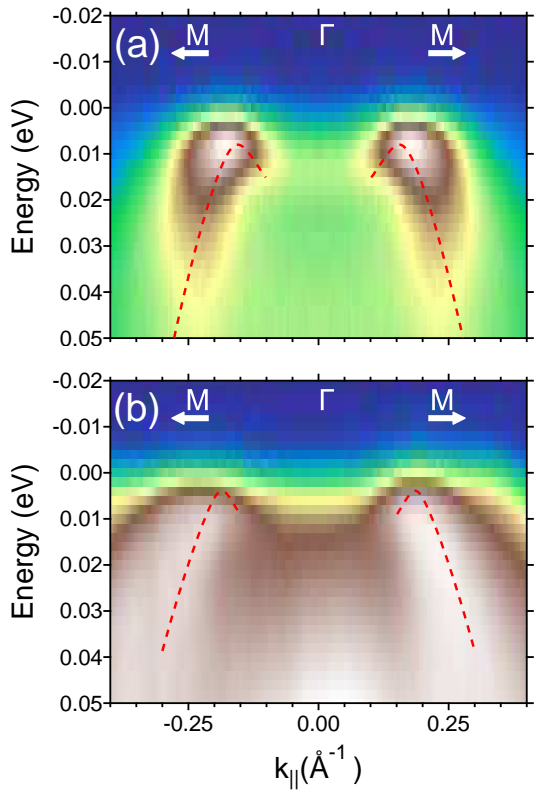}
\caption{ARPES energy-momentum maps of EuRbFe$_4$As$_4$ measured along the $\Gamma-M$ direction at a temperature of $1.5$~K using photons with an energy of $28$~eV. The intensity scale is the same as in Fig.~\ref{Fermi}. The red dashed lines are guides to the eye derived from a parabolic dispersion together with a Bogoliubov-like back dispersion at low energies. (a) Inner hole pocket measured with vertically polarized photons. (b) Middle hole pocket measured with horizontally polarized photons.}
\label{ekmap}
\end{figure}

Cuts measured with photons with h$\nu=28$~eV are depicted in Fig.~\ref{ekmap}. Near $\Gamma$ along the $\Gamma$-M direction, corresponding  to the vertical $k_y$ axis in Fig.~\ref{Fermi}, the dispersion of the  hole pockets is visible. The map in Fig.~\ref{ekmap}(a) was measured using vertically polarized photons recording spectral weight with predominantly Fe~3d$(yz)$ character. Because of matrix element effects, only the inner hole pocket is visible. Using horizontally polarized photons the spectral weight with predominantly Fe~3d$(xz)$ character of the inner hole pocket is detected (see Fig.~\ref{ekmap}(b)). Some intensity of an additional band appears above $20$~meV in the center of the BZ.

\begin{figure}
\includegraphics [width=7cm]{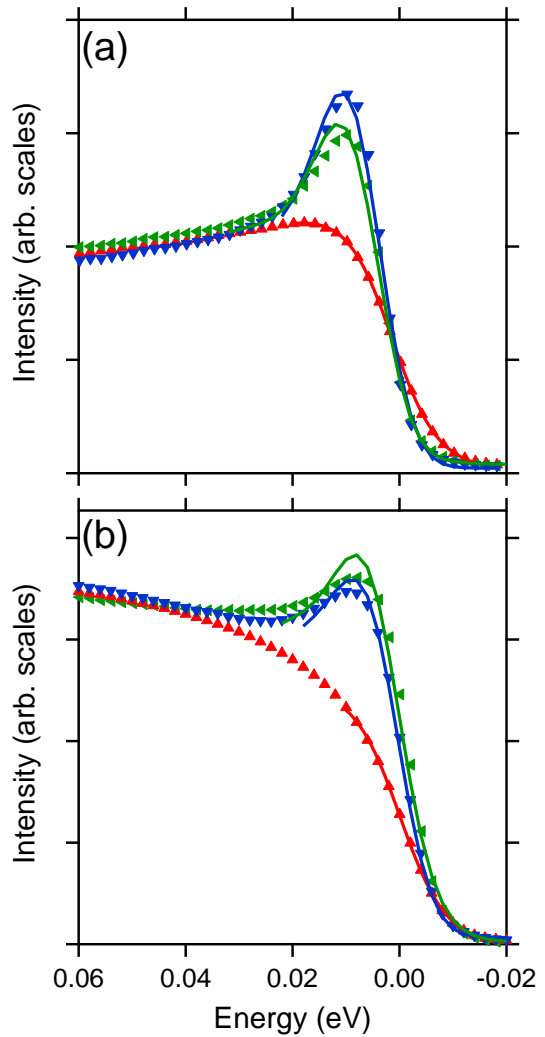}
\caption{ARPES density of states in EuRbFe$_4$As$_4$ derived from a k summation of data similar to those presented in Fig.~\ref{ekmap}. Data measured at temperatures $1.5$~K, $20$~K, and $50$~K are marked by blue, green, and red symbols, respectively. Solid lines are results from fits. For the normal state data measured at $50$~K a Fermi edge is used. For the data  measured in the superconducting state a Dynes function is used for the fit of the data. (a) Data from the inner hole pocket. (b) data from the outer hole pocket.}
\label{EDChp}
\end{figure}

In Fig.~\ref{EDChp} we present the density of states $\rho(E)$, derived from the $k$ summation of the cuts shown in Fig.~\ref{ekmap} over a range of $k_F \pm$ 0.05 {\AA}$^{-1}$ . Fig.~\ref{EDChp}(a) shows data of the inner hole pocket, while Fig.~\ref{EDChp}(b) shows data from the middle hole pocket. In both panels data for the temperatures 1.5, 20, and 50 K are presented. Normal state data are fitted with a Fermi function. The superconducting gaps $\Delta$ are derived from data measured in the superconducting state by fitting  with a Dynes function~\cite{Dynes1978}:
\begin{eqnarray}
\rho(E)=\Re \frac{E-i\Gamma_S}{((E-i\Gamma_S)^2-\Delta^2)^{0.5}}.
\label{Den}
\end{eqnarray}
Here $\Gamma_S$ is the finite width caused by the imaginary part of the order parameter. Furthermore,  we convoluted the Dynes function with a Gaussian, the width of which is determined by the finite energy resolution. For the inner hole pocket we obtain $\Delta$ values of about $8$~meV, while for the middle hole pocket we obtain values about 4 meV. Slightly higher values for the gap of the inner and the middle hole pocket were reported in Ref.~\onlinecite{Kim2020}. Considerably higher gap values, but with the same difference, have been derived  for Ba$_{0.6}$K$_{0.4}$Fe$_2$As$_2$.\cite{Nakayama2009} For $\Gamma_S$ we receive  values around $0.06$~meV typical of strong coupling superconductors.\cite{Dynes1978}

\begin{figure}
\centering
\includegraphics [width=7cm]{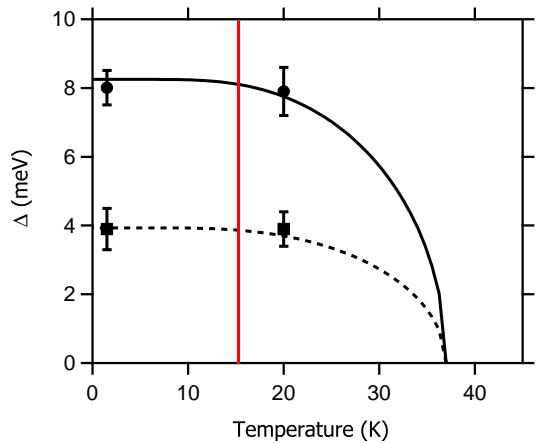}
\caption{Experimental results of the superconducting gap as a function of temperature derived from  the fits presented in Fig.~\ref{EDChp}  using data measured along the $k_y$ direction. Filled circles: data from the inner hole pocket. Squares: data from the middle hole pocket. The solid and the dashed lines present the temperature dependence expected from the weak coupling BCS theory using a superconducting transition temperature $T_{\rm c}$=36.5~K. The red line marks the ferromagnetic transition temperature $T_{\rm m}$=15~K of the Eu$^{2+}$ system.}
\label{DvsT}
\end{figure}

The central ARPES result is presented in Fig.~\ref{DvsT}, where we show the temperature dependence of the superconducting gap $\Delta$ for the inner and the middle hole pocket above and below the ferromagnetic transition temperature $T_{\rm m} = 15$~K of the Eu$^{2+}$ ions. In this figure, we have also added the temperature dependence of the superconducting gap expected within the weak-coupling BCS theory.\cite{Muehlschlegel1959} Within error bars no change of the superconducting order parameter is observed between $1.5$~K and $20$~K. This means that the order parameter of the superconducting phase does not change when the magnetic order of the Eu$^{2+}$ ions sets in.

Looking in Fig.~\ref{ekmap}(a) to the dispersion of the inner hole pocket, a strong decay of the spectral weight is detected with increasing energy. This indicates a rapid increase of the scattering rate $\Gamma(E)$ as a function of energy due to strong correlation effects. These are probably caused by a coupling between the hole and the electron pockets via spin fluctuation excitations, which is the most popular model for $s^\pm$ superconductivity for iron based superconductivity.\cite{Mazin2005} A considerably less dramatic reduction of the width at constant energy is realized for the middle hole pocket. A preliminary evaluation of the widths at constant energy multiplied with the velocity yields a ratio of the slopes of the scattering rates of the two pockets of about two.\cite{Valla1999} A more detailed analysis of the scattering results will be presented in a forthcoming paper.

The strong difference of the scattering rates of the inner and the middle hole pocket, having  Fe 3d ($yz$) and $(xz)$ orbital character along the $k_y$ axis, respectively,
is a remarkable result. It indicates that these scattering processes, which probably mediate superconductivity, are related to the symmetry of sections of the electron pocket which have the same orbital character. The difference in the scattering rates has been observed in several other iron based superconductors such as NaFeAs~\cite{Fink2015}, LiFeAs~\cite{Fink2019}, and electron and hole doped BaFe$_2$As$_2$.\cite{Fink2017,Fink2020} This observation was predicted from RPA calculation, which pointed out that intra-orbital scattering rates are larger than inter-orbital scattering rates. Because the sections having a specific orbital character in the two hole pockets are rotated by $90^{\circ}$, the coupling of the inner hole pocket to the electron pockets is considerably larger than that of the middle hole pocket.\cite{Graser2010} However this difference is in strong contrast to combined density functional plus dynamical mean field theory (DFT+DMFT) calculations, which do not show any differences between the scattering rates for these two pockets.\cite{Fink2019} As mentioned above, in the standard model for $s^\pm$ superconductivity in iron based superconductors the strength of the superconducting order parameter should be related to the strength of the inter-hole scattering rates. In this way, we can explain the larger superconducting gap for the inner hole pocket when compared to the middle hole pocket (see Fig.~\ref{DvsT}).

As discussed already above the central result of the ARPES measurements is that the superconducting order parameter does not change when the magnetic order of the Eu$^{2+}$ system appears. This shows the unique behavior of EuRbFe$_4$As$_4$ due to the strong decoupling of the magnetic and superconducting sublattices.

\section{Summary}

We have successfully synthesized high-quality single crystals of EuRbFe$_4$As$_4$. We have performed high-field magneto-transport, ESR and ARPES measurements in order to understand the interplay between the topological magnetic order of localized Eu$^{2+}$ ions and nearly isotropic superconductivity of the itinerant electrons of the Fe $3d$ band. 

ESR results for both in-plane and out-of-plane exhibit a reduced density of states on the Fermi level compared to a typical metal. It means that less amount of conduction electrons of FeAs-layers are scattered by localized magnetic moments of Eu$^{2+}$. Previous ESR study of EuFe$_{2}$As$_{2}$ also pointed out that the density of conduction electrons is significantly reduced in the SDW ground state.\cite{Dengler2010} On the other hand, vortex dynamics of Eu$^{2+}$ moments exists only for $H || c$ and is completely absent for $H || ab$, although Eu$^{2+}$ magnetic moments favor to align within the $ab$-plane. As a result, BKT phase transition is suppressed in the $ab$-plane by the strong ferromagnetism. It implies that weak ferromagnetism is required to realize a BKT phase transition. 

The anisotropy in the upper critical field of EuRbFe$_4$As$_4$ is very small at low temperature. It reflects the three-dimensional character of the Fermi surface as expected for this class of materials. ARPES measurements shows that in the presence of magnetic order on the Eu-site, the superconducting order parameter does not change. It implies that the strong decoupling of magnetic and superconducting sublattices. 

As a final conclusion, the analysis of all experimental data of this peculiar system demonstrates that superconductivity is decoupled from the Eu$^{2+}$ magnetic moments. This seems to be a direct result of the topological protection of the Eu$^{2+}$ magnetic order from conduction electrons of the FeAs-layers.

\section{Acknowledgement}

This work was supported by the German Research Foundation Project No. 477107745057 (TRR80) and by the Freistaat Bayern through the Programm f\"ur Chancengleichheit f\"ur Frauen in Forschung und Lehre. M.~H. and H.-A. K.v.N. acknowledge funding within the joint RFBR-DFG research project contract No. 19-51-45001 and KR2254/3-1. We acknowledge the support of the HLD at HZDR, member of the European Magnetic Field Laboratory (EMFL).



\begin{thebibliography}{99}


\bibitem{Ginzburg1956} V. Ginzburg, Zh. Eksp. Teor. Fiz. {\bf 31}, 202 (1956) [Sov. Phys. JETP {\bf 4}, 153 (1956)].

\bibitem{Matthais1958} B. T. Matthais, H. Suhl, and Corenzwit, Phys. Rev. Lett. {\bf 1}, 92 (1958).

\bibitem{Anderson1959} P. W. Anderson and H. Suhl, Phys. Rev. {\bf 116}, 898 (1959).

\bibitem{Fertig1977} W. A. Fertig, D. C. Johnston, L. E. DeLong, R. W. McCallum, M. B. Maple, and B. T. Matthias, Phys. Rev. Lett. \textbf{38}, 987 (1977).

\bibitem{Ishikawa1977} M. Ishikawa and O. Fischer, Solid State Commun. {\bf 23}, 37 (1977)

\bibitem{Felner1997} I. Felner, U. Asaf, Y. Levi, and O. Millo, Phys. Rev. B {\bf 55}, R3374(R) (1997).

\bibitem{Bernhard1999} C. Bernhard, J. L. Tallon, Ch. Niedermayer, Th. Blasius, A. Golnik, E. Br\"ucher, R. K. Kremer, D. R. Noakes, C. E. Stronach, and E. J. Ansaldo
Phys. Rev. B {\bf 59}, 14099 (1999).

\bibitem{Saxena2000} S. S. Saxena, P. Agarwal, K. Ahilan, F. M. Grosche, R. K. W. Haselwimmer, M. J. Steiner, E. Pugh, I. R. Walker, S. R. Julian, P. Monthoux, G. G. Lonzarich, A. Huxley, I. Sheikin, D. Braithwaite, and J. Flouquet, Nature {\bf 406}, 587 (2000).

\bibitem{Aoki2001} D. Aoki, A. Huxley, E. Ressouche, D. Braithwaite, J. Flouquet, J.-P. Brison, E. Lhotel, and C. Paulsen, Nature {\bf 413}, 613 (2001).

\bibitem{Buzdin2005} A. I. Buzdin, Rev. Mod. Phys. {\bf 77}, 935 (2005).

\bibitem{Larkin1964} A. I. Larkin, Y. N. Ovchinnikov, Zh. Eksp. Teor. Fiz. {\bf 47}, 1136 (1964) [Sov. Phys. JETP {\bf 20}, 762 (1965)].

\bibitem{Fulde1964} P. Fulde, R. A. Ferrell, Phys. Rev. {\bf 135}, A550 (1964).

\bibitem{Zdravkov2006} V. Zdravkov, A. Sidorenko, G. Obermeier, S. Gsell, M. Schreck, C. M\"uller, S. Horn, R. Tidecks, and L. R. Tagirov, Phys. Rev. Lett. {\bf 97}, 057004 (2006).

\bibitem{Lenk2016} D. Lenk, M. Hemmida, R. Morari, V. I. Zdravkov, A. Ullrich, C. M\"uller, A. S. Sidorenko, S. Horn, L. R. Tagirov, A. Loidl, H.-A. Krug von Nidda, and R. Tidecks
Phys. Rev. B {\bf 93}, 184501 (2016).

\bibitem{Kamihara2008} Y. Kamihara, T. Watanabe, M. Hirano, and H. Hosono, J. Am. Chem. Soc. {\bf 130}, 3296 (2008).

\bibitem{Saha2009} S. R. Saha, N. P. Butch, K. Kirshenbaum, J. Paglione, and P. Y. Zavalij, Phys. Rev. Lett. {\bf 103}, 037005 (2009).

\bibitem{Cao2010} G.-H. Cao, Z. Ma, C. Wang, Y. Sun, J. Bao, S. Jiang, Y. Luo, C. Feng, Y. Zhou, Z. Xie, F. Hu, S. Wei, I. Nowik, I. Felner, L. Zhang, Z.-A. Xu, and F.-C. Zhang, Phys. Rev. B {\bf 82}, 104518 (2010).

\bibitem{Luo2011} Y. K. Luo, H. Han, S. Jiang, X. Lin, Y. Li, J. Dai, G.-H. Cao, and Z.-A. Xu, Phys. Rev. B {\bf 83}, 054501 (2011).

\bibitem{Xiao2009} Y. Xiao, Y. Su, M. Meven, R. Mittal, C. M. N. Kumar, T. Chatterji, S. Price, J. Persson, N. Kumar, S. K. Dhar, A. Thamizhavel, and Th. Brueckel, Phys. Rev. B {\bf 80}, 174424 (2009).

\bibitem{Jeevan2008} H. S. Jeevan, Z. Hossain, Deepa Kasinathan, H. Rosner, C. Geibel, and P. Gegenwart, Phys. Rev. B {\bf 78}, 052502 (2008).

\bibitem{Jiao2011} W. H. Jiao, W.-H. Jiao, Q. Tao, J.-K. Bao, Y.-L. Sun, C.-M. Feng, Z.-A. Xu, I. Nowik, I. Felner and G.-H. Cao, EPL {\bf 95}, 67007 (2011).

\bibitem{Jiao2012} W. H. Jiao, I. Felner, I. Nowik and G.-H. Cao, J. Supercond Nov. Magn. {\bf 25}, 41 (2012).

\bibitem{Ren2009a} Z. Ren, X. Lin, Q. Tao, S. Jiang, Z. Zhu, C. Wang, G.-H. Cao, and Z.-A. Xu, Phys. Rev. B {\bf 79}, 094426 (2009).

\bibitem{Nandi2014} S. Nandi, W. T. Jin, Y. Xiao, Y. Su, S. Price, W. Schmidt, K. Schmalzl, T. Chatterji, H. S. Jeevan, P. Gegenwart, and Th. Br\"uckel, Phys. Rev. B {\bf 90}, 094407 (2014).

\bibitem{Ren2009} Z. Ren, Q. Tao, S. Jiang, C. Feng, C. Wang, J. Dai, G.-H. Cao, and Z. Xu, Phys. Rev. Lett. {\bf 102}, 137002 (2009).

\bibitem{Jiang2009} S. Jiang, H. Xing, G. Xuan, Z. Ren, C. Wang, Z.-A. Xu, and G.-H. Cao, Phys. Rev. B {\bf 80}, 184514 (2009).

\bibitem{Ahmed2010} A. Ahmed, M. Itou, S. Xu, Z.-A. Xu, G.-H. Cao, Y. Sakurai, J. Penner-Hahn, and A. Deb, Phys. Rev. Lett. {\bf 105}, 207003 (2010).

\bibitem{Nowik2011}  I Nowik, I Felner, Z. Ren, G.-H. Cao and Z. A. Xu, J. Phys. Condens. Matter {\bf 23}, 065701 (2011).

\bibitem{Wu2011} D. Wu, G. Chanda, H. S. Jeevan, P. Gegenwart, and M. Dressel, Phys. Rev. B {\bf 83}, 100503(R) (2011).

\bibitem{Jeevan2011} H. S. Jeevan, D. Kasinathan, H. Rosner, and P. Gegenwart, Phys. Rev. B {\bf 83}, 054511 (2011).

\bibitem{Zapf2011} S. Zapf, D. Wu, L. Bogani, H. S. Jeevan, P. Gegenwart, and M. Dressel, Phys. Rev. B {\bf 84}, 140503(R) (2011).

\bibitem{Munevar2014} J. Munevar, H. Micklitz, M. Alzamora, C. Arg\"ullo, T. Goko, F. L. Ning, A. A. Aczel, T. Munsie, T. J. Williams, G. F. Chen, W. Yu, G. M. Luke, Y. J. Uemura, and E. Baggio-Saitovitch, Solid State Commun. {\bf 187}, 18 (2014).

\bibitem{Hemmida2014} M. Hemmida, H.-A. Krug von Nidda, A. G\"unther, A. Loidl, A. Leithe-Jasper, W. Schnelle, H. Rosner, and J. Sichelschmidt, Phys. Rev. B {\bf 90}, 205105 (2014). 

\bibitem{Iyo2016} A. Iyo, K. Kawashima, T. Kinjo, T. Nishio, S. Ishida, H. Fujihisa, Y. Gotoh, K. Kihou, H. Eisaki, and Y. Yoshida, J. Am. Chem. Soc. {\bf 138}, 3410 (2016). 

\bibitem{Kawashima2016} K. Kawashia, T. Kinjo, T. Nishio, S. Ishida, H. Fujihisa, Y. Gotoh, K. Kihou, H. Eisaki, Y. Yoshida, and A. Iyo, J. Phys. Soc. Jpn. {\bf 85}, 064710 (2016).

\bibitem{Liu2016} Y. Liu, Y.-B. Liu, Z.-T. Tang, H. Jiang, Z.-C. Wang, A. Ablimit, W.-H. Jiao, Q. Tao, C.-M. Feng, Z.-A. Xu, and G.-H. Cao, Phys. Rev. B {\bf 93}, 214503 (2016).

\bibitem{Wang2017} Z. Wang, C. He, Z. Tang, S. Wu, and G.-H. Cao, Sci. China Mater {\bf 60}, 83 (2017).

\bibitem{Bao2018} J.-K. Bao, K. Willa, M. P. Smylie, H. Chen, U. Welp, D. Y. Chung, and M. G. Kanatzidis, Cryst. Growth Des. {\bf 18}, 3517 (2018).

\bibitem{Iida2019} K. Iida, Y. Nagai, S. Ishida, M. Ishikado, N. Murai, A. D. Christianson, H. Yoshida, Y. Inamura, H. Nakamura, A. Nakao, K. Munakata, D. Kagerbauer, M. Eisterer, K. Kawashima, Y. Yoshida, H. Eisaki, and A. Iyo, Phys. Rev. B {\bf 100}, 014506 (2019).

\bibitem{Devizorova2019a} Zh. Devizorova, S. Mironov, and A. Buzdin, Phys. Rev. Lett. {\bf 122}, 117002 (2019).

\bibitem{Devizorova2019b} Zh. Devizorova, and A. Buzdin, Phys. Rev. B {\bf 100}, 104523 (2019).

\bibitem{Koshelev2019a} A. E. Koshelev, Phys. Rev. B {\bf 100}, 224503 (2019).

\bibitem{Xu2019} C. Xu, Q. Chen, and C. Cao, Commun. Phys. {\bf 2}, 16 (2019).

\bibitem{Nejadsattari2020} F. Nejadsattari, M. A. Albedah, and Z. M. Stadnik, Phil. Mag. {\bf 100}, 894 (2020).

\bibitem{Liu2017} Y. Liu, Y.-B. Liu, Y.-L. Yu, Q. Tao, C.-M. Feng, and G.-H. Cao, Phys. Rev. B {\bf 96}, 224510 (2017).

\bibitem{Jackson2018} D. E. Jackson, D. VanGennep, W. Bi, D. Zhang, P. Materne, Y. Liu, G.-H. Cao, S. T. Weir, Y. K. Vohra, and J. J. Hamlin, Phys. Rev. B {\bf 98}, 014518 (2018).

\bibitem{Smylie2018} M. P. Smylie, K. Willa, J.-K. Bao, K. Ryan, Z. Islam, H. Claus, Y. Simsek, Z. Diao, A. Rydh, A. E. Koshelev, W.-K. Kwok, D. Y. Chung, M. G. Kanatzidis, and U. Welp, Phys. Rev. B {\bf 98}, 104503 (2018).

\bibitem{Stolyarov2018} V. S. Stolyarov, A. Casano, M. A. Belyanchikov, A. S. Astrakhantseva, S. Y. Grebenchuk, D. S. Baranov, I. A. Golovchanskiy, I. Voloshenko, E. S. Zhukova, B. P. Gorshunov, A. V. Muratov, V. V. Dremov, L. Y. Vinnikov, D. Roditchev, Y. Liu, G.-H. Cao, M. Dressel, and E. Uykur, Phys. Rev. B {\bf 98}, 140506(R) (2018).

\bibitem{Smylie2019} M. P. Smylie, A. E. Koshelev, K. Willa, R. Willa, W.-K. Kwok, J.-K. Bao, D. Y. Chung, M. G. Kanatzidis, J. Singleton, F. F. Balakirev, H. Hebbeker, P. Niraula, E. Bokari, A. Kayani, and U. Welp, Phys. Rev. B {\bf 100}, 054507 (2019).

\bibitem{Xiang2019} L. Xiang, S. L. Bud'ko, J.-K. Bao, D. Y. Chung, M. G. Kanatzidis, and P. C. Canfield, Phys. Rev. B {\bf 99}, 144509 (2019).

\bibitem{Willa2019} K. Willa, R. Willa, J.-K. Bao, A. E. Koshelev, D. Y. Chung, M. G. Kanatzidis, W.-K. Kwok, and U. Welp, Phys. Rev. B {\bf 99}, 180502(R) (2019).

\bibitem{Koshelev2019} A. E. Koshelev, K. Willa, R. Willa, M. P. Smylie, J.-K. Bao, D. Y. Chung, M. G. Kanatzidis, W.-K. Kwok, and U. Welp, Phys. Rev. B {\bf 100}, 094518 (2019).

\bibitem{Holenstein2019} S. Holenstein, B. Fischer, Y. Liu, N. Barbero, G. Simutis, Z. Shermadini, M. Elender, P. K. Biswas, R. Khasanov, A. Amato, T. Shiroka, H.-H. Klauss, E. Morenzoni, G.-H. Cao, D. Johrendt, H. Luetkens, arXiv:1911.04325 (2019).

\bibitem{Liu2020} Y.-B. Liu, Y. Liu, Y.-W. Cui, Z. Ren, and G.-H. Cao, J. Phys. Condens. Matter {\bf 32}, 175701 (2020).

\bibitem{Willa2020} K. Willa, M. P. Smylie, Y. Simsek, J.-K. Bao, D. Y. Chung, M. G. Kanatzidis, W.-K. Kwok, and U. Welp, Phys. Rev. B {\bf 101} 064508 (2020).

\bibitem{Vlasov2020} V. K. Vlasko-Vlasov, U. Welp, A. E. Koshelev, M. Smylie, J.-K. Bao, D. Y. Chung, M. G. Kanatzidis, and W.-K. Kwok, Phys. Rev. B {\bf 101}, 104504 (2020).

\bibitem{Vlasenko2020} V.A. Vlasenko, K.S. Pervakov, S.U. Gavrilkin, arXiv:1906.04597 (2020).

\bibitem{Kim2020} T. K. Kim, K. S. Pervakov, D. V. Evtushinsky, S. W. Jung, G. Poelchen, K. Kummer, V. A. Vlasenko, V. M. Pudalov, D. Roditchev, V. S. Stolyarov, D. V. Vyalikh,
V. Borisov, R. Valentı, A. Ernst, S. V. Eremeev, and E. V. Chulkov, arXiv:2008.00736 (2020).

\bibitem{Meier2017} W. R. Meier, T. Kong, S. L. Bud'ko, and P. C. Canfield, Phys. Rev. Materials {\bf 1}, 013401 (2017).

\bibitem{Fink2009} J. Fink, S. Thirupathaiah, R. Ovsyannikov, H. A. Duerr,  R. Follath, Y. Huang. S. de Jong, M. S. Golden, Y.-Z. Zhang, H. O. Jeschke, R. Valenti, C. Felser, 
S.~D. Farahani, M. Rotter, and D. Johrendt, Phys. Rev. B {\bf 79}, 155118 (2009).

\bibitem{Moser2017} S. Moser, Journal of Electron Spectroscopy and Related Phenomena {\bf 214}, 29 (2017).

\bibitem{Barnes1981} S. E. Barnes, Adv. Phys. {\bf 30}, 801 (1981).

\bibitem{Owens2001} F. J. Owens, physica C {\bf 353}, 265 (2001).

\bibitem{Joshi2004} J. p. Joshi and S. V. Bhat, J. Magn. Reson. {\bf 168}, 284 (2004).

\bibitem{Korringa1950} J. Korringa, Physica {\bf 16}, 601 (1950).

\bibitem{Dengler2010} E. Dengler, J. Deisenhofer, H.-A. Krug von Nidda, S. Khim, J. S. Kim, K. H. Kim, F. Casper, C. Felser, and A. Loidl, Phys. Rev. B {\bf 81}, 024406 (2010).

\bibitem{Pascher2010} N. Pascher, J. Deisenhofer, H.-A. Krug von Nidda, M. Hemmida, H. S. Jeevan, P. Gegenwart, and A. Loidl, Phys. Rev. B {\bf 82}, 054525 (2010).

\bibitem{Krug2012} H.-A. Krug von Nidda, S. Kraus, S. Schaile, E. Dengler, N. Pascher, M. Hemmida, M. J. Eom, J. S. Kim, H. S. Jeevan, P. Gegenwart, J. Deisenhofer, and A. Loidl, Phys. Rev. B {\bf 86}, 094411 (2012).

\bibitem{Taylor1975} R. H. Taylor, Adv. Phys. {\bf 24}, 681 (1975).

\bibitem{Berezinskii1972} V. Berezinskii, Zh. Eksp. Teor. Fiz {\bf 61}, 610 (1972).

\bibitem{Kosterlitz1973} J. M. Kosterlitz and D. J. Thouless, J. Phys. C {\bf 6}, 1181 (1973).

\bibitem{Kosterlitz1974} J. M. Kosterlitz, J. Phys. C {\bf 7}, 1046 (1974).

\bibitem{Heinrich2003} M. Heinrich, H.-A. Krug von Nidda, A. Loidl, N. Rogado, R. J. Cava, Phys. Rev. Lett. {\bf 91}, 137601 (2003).

\bibitem{Demokritov1989} S. O. Demokritov, M. M. Kreines, V. I. Kudinov, and S. V. Detra, Zh. Eksp. Teor. Fiz. {\bf 95}, 2211 (1989)[Sov. Phys. JETP {\bf 68}, 1277 (1989)].

\bibitem{Gaveau1991} P. Gaveau, J. P. Boucher, L. P. Regnault, and Y. Henry, J. Appl. Phys. {\bf 69}, 6228 (1991).

\bibitem{Golubov2011} A. A. Golubova, O. V. Dolgovb, A. V. Borisb, A. Charnukhab, D. L. Sunb, C. T. Linb, A. F. Shevchunc, A. V. Korobenkoc, M. R. Truninc, and V. N. Zverev, JETP Letters {\bf 94}, 333 (2011).

\bibitem{Shen2011} B. Shen, H. Yang, Z.-S. Wang, F. Han, B. Zeng, L. Shan, C. Ren, and H.-H. Wen, Phys. Rev. B {\bf 84}, 184512 (2011).

\bibitem{Kurita2011} N. Kurita, M. Kimata, K. Kodama, A. Harada, M. Tomita, H. S. Suzuki, T. Matsumoto, K. Murata, S. Uji, and T. Terashima, Phys. Rev. B {\bf 83}, 214513 (2011).

\bibitem{Paramanik2013} U. B. Paramanik, D. Das, R Prasad and Z Hossain, J. Phys.: Condens. Matter {\bf 25 }, 265701 (2013).

\bibitem{Jiang2009a} S. Jiang, Y. Luo, Z. Ren, Z. Zhu, C. Wang, X. Xu, Q. Tao, G.-H. Cao and Z. Xu, New Journal of Physics {\bf 11}, 025007 (2009).

\bibitem{Terashima2010} T. Terashima, N. Kurita, A. Kikkawa, H. S. Suzuki, T. Matsumoto, K. Murata, and S. Uji, J. Phys. Soc. Jpn. {\bf 79}, 103706 (2010).

\bibitem{Werthamer1966} N. Werthamer, E. Helfand, P. Hohenberg, Phys. Rev. {\bf 147}, 295 (1966).

\bibitem{Tinkham1996} M. Tinkham, \textit{Introduction to Superconductivity}, 2nd edition (McGraw-Hill, New York, 1996).

\bibitem{Tuyn1925} W. Tuyn and H. Kamerlingh Onnes, Leiden Comm. {\bf 174a} (1925).

\bibitem{Meier2016} W. R. Meier, T. Kong, U. S. Kaluarachchi, V. Taufour, N. H. Jo, G. Drachuck, A. E. B\"ohmer, S. M. Saunders, A. Sapkota, A. Kreyssig, M. A. Tanatar, R. Prozorov, A. I. Goldman, Fedor F. Balakirev, Alex Gurevich, S. L. Bud'ko, and P. C. Canfield, Phys. Rev. B {\bf 94}, 064501 (2016).

\bibitem{Izyumov2010} Y. Izyumov and E. Kurmaev, \textit{High-T$_{\rm c}$ Superconductors Based on FeAs Compounds}, 1st edition, Springer Series in Material Science, Vol. 143 (Springer-Verlag, Berlin-Heidelberg, 2010).

\bibitem{Yuan2009} H.Q. Yuan, J. Singleton, F.F. Balakirev, S.A. Baily, G.F. Chen, J.L. Luo, N.L. Wang, Nature {\bf 457}, 565 (2009). 

\bibitem{Zhang2011} J.-L. Zhang, L. Jiao, Y. Chen, H.-Q. Yuan, Front. Phys. {\bf 6}, 463 (2011).

\bibitem{Sebastian2008} S. E. Sebastian, N. Harrison, E. Palm, T. P. Murphy, C. H. Mielke, R. Liang, D. A. Bonn, W. N. Hardy, and G. G. Lonzarich, Nature (London) {\bf 454}, 200 (2008).

\bibitem{Goddard2004} P. A. Goddard, S. J. Blundell, J. Singleton, R. D. McDonald, A. Ardavan, A. Narduzzo, J. A. Schlueter, A. M. Kini, and T. Sasaki, Phys. Rev. B {\bf 69}, 174509 (2004).

\bibitem{Lei2010} H. Lei, R. Hu, E. S. Choi, J. B. Warren, and C. Petrovic, Phys. Rev. B {\bf 81}, 184522 (2010).

\bibitem{Fang2010} M. Fang, J. Yang, F. F. Balakirev, Y. Kohama, J. Singleton, B. Qian, Z. Q. Mao, H. Wang, and H. Q. Yuan, Phys. Rev. B {\bf 81}, 020509(R) (2010).

\bibitem{Khim2010} S. Khim, J. W. Kim, E. S. Choi, Y. Bang, M. Nohara, H. Takagi, and K. H. Kim, Phys. Rev. B {\bf 81}, 184511 (2010). 

\bibitem{Chandrasekhar1962} B. S. Chandrasekhar, Appl. Phys. Lett. {\bf 1}, 7 (1962).

\bibitem{Clogston1962} A. M. Clogston, Phys. Rev. Lett. {\bf 9}, 266 (1962).

\bibitem{Maki1966} K. Maki, Phys. Rev. {\bf 148}, 362 (1966).

\bibitem{Kida2009} T. Kida, T. Matsunaga, M. Hagiwara, Y. Mizuguchi, Y. Takano, and K. Kindo, J. Phys. Soc. Jpn. {\bf 78}, 113701 (2009).

\bibitem{Damascelli2003} A. Damascelli, Z. Hussain, and Z.-X. Shen, Rev. Mod. Phys. {\bf 75}, 473 (2003).
	
\bibitem{Dynes1978} R. C. Dynes, V. Narayanamurti,  and J. P. Garno,  Phys. Rev. Lett. {\bf 41}, 1509 (1978).

\bibitem{Nakayama2009} K. Nakayama, T. Sato, P. Richard, Y.-M. Xu, Y. Sekiba, S. Souma, G. F. Chen, J. L. Luo, N. L. Wang, H. Ding and T. Takahashi, EPL {\bf 85}, 67002 (2009). 

\bibitem{Muehlschlegel1959}  H. Muehlschlegel, Z. Phys. {\bf 155}, 313 (1957).

\bibitem{Mazin2005} I. I. Mazin, D. J. Singh, M. D.  Johannes,  and M. H. Du, Phys. Rev. Lett. {\bf 101}, 057003 (2005).

\bibitem{Valla1999} T. Valla, A. V. Fedorov, P. D. Johnson, and S. L. Hulbert, Phys. Rev. Lett. {\bf 83}, 2085 (1999).
	
\bibitem {Fink2015} J. Fink, A. Charnukha, E.~D.~L. Rienks, Z. H. Liu, S. Thirupathaiah, I. Avigo, F. Roth, H.~S. Jeevan, P. Gegenwart, M. Roslova, I. Morozov, S. Wurmehl, U. Bovensiepen, S. Borisenko, M. Vojta, and B. B\"uchner, Phys. Rev. B {\bf 92}, 201106 (2015).

\bibitem {Fink2019} J. Fink, J. Nayak, E. D. L. Rienks, J. Bannies, S. Wurmehl, S. Aswartham, I. Morozov, R. Kappenberger, M. A. ElGhazali, L. Craco, H. Rosner, C. Felser, and B. B\"uchner, Phys. Rev. B {\bf 99}, 245156 (2019).
	
\bibitem {Fink2017} J. Fink, E.~D.~L. Rienks, S. Thirupathaiah, J. Nayak, A.van Roekeghem,S. Biermann,T. Wolf, P. Adelmann, H.~S. Jeevan, P.~Gegenwart, S. Wurmehl, C. Felser, and B. B\"uchner, Phys. Rev. B {\bf 95}, 144513 (2017).
	
\bibitem {Fink2020} J. Fink, E.D.L. Rienks, M.Yao, R. Kurleto, J. Bannies, S. Aswartham, I. Morozov, S.Wurmehl, T.Wolf, F. Hardy, C. Meingast, H.S. Jeevan, J. Maiwald,,  P. Gegenwart, C. Felser, and B. B\"uchner, arXiv:2005.08216 (2020).

\bibitem {Graser2010} S. Graser, A. F. Kemper, T. A. Maier,  H.-P. Cheng,  P. J. Hirschfeld,  and D. J. Scalapino, Phys. Rev. B {\bf 81} 214503 (2010).



\end{thebibliography}
\end{document}